\documentstyle[aps]{revtex}

\begin{document}
\title{Enhanced the thermal Entanglement in Anisotropy Heisenberg $XYZ$ Chain}
\author{L. Zhou, H. S. Song, Y. Q. Guo and C. Li}
\address{{\small Department of Physics, Dalian University of Technology,}\\
Dalian, 116024, P. R. China}
\maketitle

\begin{abstract}
The thermal entanglement in Heisenberg $XYZ$ chain is investigated in the
presence of external magnetic field $B$. In the two-qubit system, the
critical magnetic field $B_c$ is increased because of introducing the
interaction of the z-component of two neighboring spins $J_z$. This
interaction not only improves the critical temperature $T_c,$ but also
enhances the entanglement for particular fixed $B$. We also analyze the
pairwise entanglement between nearest neighbors in three qubits. The
pairwise entanglement, for a fixed $T$, can be strong by controlling $B$ and 
$J_z$.

\end{abstract}

\section{Introduction}

\bigskip Entanglement is an important resource in quantum information \cite
{bennett}. The ideal case which Quantum computing and quantum communication
\ are put into use \ is to find entanglement resource in solid system at a
finite temperature. The Heisenberg model is a simple but realistic and
extensively studied solid state system \cite{PRH}\cite{SE}. Recently,
Heisenberg interaction is not localized in spin system. It can be realized
in quantum dots \cite{DL}, nuclear spins \cite{BEK}, cavity QED \cite{AI}%
\cite{SBZ}. This effect Hamiltonian can be used for quantum computation \cite
{DAL} and controlled NOT gate \cite{SBZ}. The thermal entanglement in
isotropic Heisenberg spin chain has been studied in the absence [9,10,15]
and in the presence of an external magnetic $B$ [9,10,14]. The entanglement
of two-qubit isotropic Heisenberg system decreases with the increasing $T$
and vanishes beyond a critical value $T_c$ [9,10], which is independent of $%
B.$ Pairwise entanglement in $N$-qubit isotropic Heisenberg system in
certain degree can be increased by increasing the temperature or the
external field $B$ \cite{MC}. Anisotropic Heisenberg spin chain has been
investigated in the case of $B=0$ \cite{wxg} and $B\neq 0$ \cite{GL}. For a
two-qubit anisotropic Heisenberg $XY$ chain, one is able to produce
entanglement for finite $T$ by adjusting the magnetic field strength \cite
{GL}. However, the entanglement by increasing $T$ or $B$, in two-qubit
anisotropic Heisenberg $XY$ chain \cite{GL} or in $N$-qubit isotropic
Heisenberg chain \cite{MC}, is very weak. How to produce strong entanglement
is worthy to study.

On the other hand, we have not find the work about two-qubit or the $N$%
-qubit anisotropic $XYZ$ Heisenberg chain in the presence of magnetic field.
Although the $N$-qubit Heisenberg chain has been studied \cite{KMO} \cite{MC}%
, in Ref. \cite{KMO} the authors studied the maximum possible nearest
neighbor entanglement for ground state in a ring of $N$ qubits, and in \cite
{MC} they just investigated the case of isotropy $N$ qubits Heisenberg
chain. In this paper, we study the entanglement of two-qubit anisotropic
Heisenberg $XYZ$ chain and the pairwise entanglement of three-qubit
anisotropic Heisenberg $XYZ$ chain. Introducing the interaction of the
z-component of two neighboring spins not only improve the critical
temperature $T_c$ but also enhance the entanglement for fixed $B$ and $T$ in
particular regions. In the case of anisotropic three-qubit Heisenberg $XYZ$
chain, the effect of partial anisotropy $\gamma $ make the revival
phenomenon more apparent than in two-qubit chain; for a fixed $T$, one can
obtain a robust entanglement by controlling $B$ and $J_z$.

The Hamiltonian of $N$-qubit anisotropic Heisenberg $XYZ$ model in an
external magnetic field $B$ is \cite{GL} 
\begin{equation}
H=\frac 12\sum_{i=1}^N[J_x\sigma _i^x\sigma _{i+1}^x+J_y\sigma _i^y\sigma
_{i+1}^y+J_z\sigma _i^z\sigma _{i+1}^z+B(\sigma _i^z+\sigma _{i+1}^z)],
\end{equation}
where $\overrightarrow{\sigma }_j=(\sigma _j^x,\sigma _j^y,\sigma _j^z)$ is
the vector of Pauli matrices and $J_i(i=x,y,z)$ is real coupling
coefficient. The coupling coefficient $J_i$ of arbitrary nearest neighbor
two qubits is equal in value. For the spin interaction, the chain is said to
be antiferromagnetic for $J_i>0$ and ferromagnetic for $J_i$ $<0$.

For a system in equilibrium at temperature $T$ , the density operator is $%
\rho =Z^{-1}\exp (-H/k_BT)$, where $Z=Tr[\exp (-H/k_BT)]$ is the partition
function and $k_B$ is Boltzmann's constant. For simplicity we write $k_B=1$.
Entanglement of two qubits can be measured by concurrence $C$ which is
written as $C=\max (0,2\max \{\lambda _i\}-\sum_{i=1}^4\lambda _i)$ \cite
{CHB}\cite{sh}\cite{ca}, where $\lambda _i$ is the square roots of the
eigenvalues of the matrix $R=\rho S\rho ^{*}S$, $\rho $ is the density
matrix, $S=\sigma _1^y\otimes \sigma _2^y$ and $*$ stand for complex
conjugate. The concurrence is available no matter what $\rho $ is pure or
mixed.

\ 

\section{Two-qubit Heisenberg $XYZ$ chain}

Now, we consider the Hamiltonian for anisotropic two-qubit Heisenberg $XYZ$
chain in an external magnetic field $B$. The Hamiltonian can be expressed as 
\begin{equation}
H=J(\sigma _1^{+}\sigma _2^{-}+\sigma _1^{-}\sigma _2^{+})+J\gamma (\sigma
_1^{+}\sigma _2^{+}+\sigma _1^{-}\sigma _2^{-})+\frac{J_z}2\sigma _1^z\sigma
_2^z+\frac B2(\sigma _1^z+\sigma _2^z)
\end{equation}
where $\sigma ^{\pm }=\frac 12(\sigma ^x\pm i\sigma ^y)$ is raising and
lowering operator respectively, and $J=\frac{J_x+J_y}2,$ $\gamma =\frac{%
J_x-J_y}{J_x+J_y}$. The parameter $\gamma $ $(0<\gamma <1)$ measure the
anisotropy (partial anisotropy) in $XY$ plane. When the Hamiltonian of the
system has the form of Eq.(2), in the standard basis $\{|00\rangle
,|01\rangle ,|10\rangle ,|11\rangle \}$, the density matrix of the system
can be written as 
\begin{equation}
\rho _{12}=\left( 
\begin{array}{llll}
u_1 & 0 & 0 & v \\ 
0 & w & z & 0 \\ 
0 & z & w & 0 \\ 
v & 0 & 0 & u_2
\end{array}
\right) .
\end{equation}
These nonzero matrix element can be calculated through 
\begin{eqnarray}
u_1 &=&Tr(|00\rangle \langle 00|\rho ),u_2=Tr(|11\rangle \langle 11|\rho ), 
\nonumber \\
w &=&Tr(|01\rangle \langle 01|\rho ),v=Tr(|00\rangle \langle 11|\rho
),z=Tr(|01\rangle \langle 10|\rho ).
\end{eqnarray}
The square roots of the eigenvalues of the matrix $R$ are $\lambda
_{1,2}=|w\pm z|,\lambda _{3,4}=|\sqrt{u_1u_2}\pm v|$. Therefore, we can
calculate the concurrence.

The eigenvalues and eigenstates of $H$ are easily obtained as $H|\Psi ^{\pm
}\rangle =(-\frac{J_z}2\pm J)|\Psi ^{\pm }\rangle $, $H|\Sigma ^{\pm
}\rangle =(\frac{J_z}2\pm \eta )|\Sigma ^{\pm }\rangle ,$with the
eigenstates $|\Psi ^{\pm }\rangle =\frac 1{\sqrt{2}}(|01\rangle \pm
|10\rangle )$, $|\Sigma ^{\pm }\rangle =\frac 1{\sqrt{2\eta (\eta \mp B)}}%
[(\eta \mp B)|00\rangle \pm J\gamma |11\rangle ]$, where $\eta =\sqrt{%
B^2+(J\gamma )^2}$. One can notice that the eigenstates are the same as the
case of $J_z=0$\cite{GL}. Because the basises $|01\rangle $ and $|10\rangle $
are the two degenerate eigenstates of $\sigma _1^z\sigma _2^z$ with
eigenvalue $-1$, hence the superposition of the two degenerate states $%
|01\rangle $ and $|10\rangle $ still is the eigenstate of $\sigma _1^z\sigma
_2^z$, that is, $|\Psi ^{\pm }\rangle $ is the eigenstate of $J_z=0$ as well
as that of $J_z\neq 0$. The same reason account for $|\Sigma ^{\pm }\rangle $
both as an eigenstate of Eq.(2) and as that of the case of $J_z=0$ . From
Eq.(4), tracing on the eigenstates, we obtain the square roots of the
eigenvalues of the matrix $R$ 
\begin{eqnarray}
\lambda _{1,2} &=&Z^{-1}e^{\frac{\beta J_z}2}e^{\pm \beta J},  \nonumber \\
\lambda _{3,4} &=&Z^{-1}e^{-\frac{\beta J_z}2}|\sqrt{1+(\frac{J\gamma }\eta
\sinh \beta \eta )^2}\mp \frac{J\gamma }\eta \sinh \beta \eta |,
\end{eqnarray}
where the partition function $Z=2(e^{-\frac{J_z}{2T}}\cosh \beta \eta +e^{%
\frac{\beta J_z}2}\cosh \beta J)$. Because the concurrence is invariant
under the substitutions $J\rightarrow -J$ and $\gamma \rightarrow -\gamma $ 
\cite{GL}, we will consider the case $J>0$ and $0<\gamma <1$. But with
substitution $J_z\rightarrow -J_z$ the the concurrence is variant. We choose 
$J_z>0$, and we will state the reason later.

We first review the circumstance of anisotropic Heisenberg $XY$ chain, which
is analyzed in \cite{GL}. At $T=0,$ exist a critical magnetic field $B_c$.
As $B$ cross $B_c$, the concurrence $C$ drops suddenly then undergoes a
''revival'' for sufficient large $\gamma .$ However, we noticed that $B_c$
decrease with the increasing of the anisotropic parameter $\gamma $.
Although with $\gamma $ increasing the critical temperature $T_c$ is
improved, the entanglement, when temperature is in the revival region, is
very weak.

With $\gamma =0.3$, we show the concurrence as a function of $B$ and $T$ for
two values of $J_z$ in Fig. 1. For $J_z=0$ (Fig.1a) corresponding to the
circumstance of anisotropic Heisenberg $XY$ chain \cite{GL}, one can observe
a revival phenomenon and the weak entanglement in revival region. For the
convenience of representation, we define the main region in which
concurrence $C$ keeping its constant and maximal value. Comparing Fig. 1 (a)
with (b), we find that with the increasing of $J_z,$ the main region is
extended in terms of $B$ and $T$, i.e., the critical magnetic field $B_c$ is
broadened and the critical temperature $T_c$ in main region is improved.
That is to say, the range of concurrence $C$ keeping its constant and
maximal is extended in terms of $B$ and $T,$ so we can obtain strong
entanglement in the extended range.

We can understand the effect of $J_z$ on $B_c$ from the case of $T=0$. For $%
T=0$ under the condition of $J_z\leq J$, $C$ can be written analytically as 
\begin{equation}
C(T=0)=\left\{ 
\begin{array}{c}
1\text{ \qquad \qquad \quad \quad for }\eta <J+J_z \\ 
(1-J\gamma /\eta )/2\text{ \quad for }\eta =J+J_z \\ 
J\gamma /\eta \text{ \quad \quad \quad \quad for }\eta >J+J_z
\end{array}
\right.
\end{equation}
The parameters $J$, $\eta $ and $\gamma $ are independent of $J_z$ in the
case of two interacting qubits. Comparing Eq.(6) with Eq.(6) of Ref.\cite{GL}%
, we can see clearly that if $J_z$ is positive, $J_z$ makes the intersection
points of piecewise function shift. In this paper, we consider the case of $%
J_z>0$. Fig. 2 shows the concurrence at $T=0$ for three values of positive $%
J_z$. It show clearly that concurrence drops sharply at a finite value of
magnetic field $B$, which is called critical magnetic field $B_c$, at which
the quantum phase transition occurs\cite{GL}. But with the increasing of $%
J_z $, $B_c$ is increased. The interaction of the z-component of two
neighboring spins $J_z$ causes a shift in the locations of the phase
transitions. Namely, the presence of positive Jz increases the region over
which the concurrence C attains its maximum value.This result means that in
larger region of $B$ and $T$ we can obtain stronger entanglement. The effect
of $J_z $ is different with that of $\gamma $ on changing $B_c$. In the case
of $J_z=0$ \cite{GL}, although with the increasing of $\gamma $ the critical
temperature $T_c$ is increased, the larger the values of $\gamma $, the
smaller the critical magnetic field $B_c$. Here, introducing the z-component
interaction of two neighboring spins not only extends critical magnetic
field $B_c$ but also improves critical temperature $T_c$ and the
entanglement (we will further show it in Fig. 3).

Let us consider concurrence changing with temperature for different values
of $J_z$ in a fixed $B$ ($B=1.1$). We plot it in Fig. 3 with $\gamma =0.3$.
We notice that existing a critical temperature $T_c$ at which the
entanglement vanishes. Obviously, $T_c$ is improved monotonously with
increasing of $J_z$. Under the condition $J_z=0$ (corresponding to $XY$
model \cite{GL}), the concurrence exhibit a revival phenomenon, but the
maximal values of entanglement in both area are small. If introducing the $%
J_z$, the critical external magnetic field $B_c$ become larger so that $%
B=1.1 $ is less than $B_c$ (the critical magnetic when $J_z=0.2,0.5$ or $%
J_z=0.9$), thus we observe the maximal value of entanglement 1. In the
temperature range $0<T<1.725$ , the larger $J_z$ the stronger entanglement.
Therefore, $J_z$ not only improve the critical temperature $T_c,$ but also
enhance the entanglement for particular fixed $B$ and $\gamma .$

\section{The pairwise entanglement in three qubits}

The calculation of pairwise entanglement in $N$ qubits is very complicated
due to the anisotropy in Heisenberg $XYZ$ chain. Here we just calculate the
pairwise entanglement in three qubits to show the effects of $J_z$ . We now
solve the eigenvalue problems of the three-qubit $XYZ$ Hamiltonian. We list
the eigenvalues and the corresponding eigenvectors as follow

\begin{eqnarray}
E_{1,2} &=&-J-\frac{J_z}2+B:|\Phi _{1,2}\rangle =\pm \frac 12(1\mp \frac 1{%
\sqrt{3}})|110\rangle +\frac 1{\sqrt{3}}|101\rangle \mp \frac 12(1\pm \frac 1%
{\sqrt{3}})|011\rangle ),  \nonumber \\
E_{3,4} &=&J+\frac{J_z}2-B\pm \eta _{-}:|\Phi _{3,4}\rangle =\frac 1{\sqrt{%
2\eta _{-}[\eta _{-}\pm (J_z-2B-J)]}}[(J_z-2B-J\pm \eta _{-})|000\rangle
+J\gamma \sum_{n=0}^2\Upsilon ^n|110\rangle ];  \nonumber \\
E_{5,6} &=&-J-\frac{J_z}2-B:|\Phi _{5,6}\rangle =\pm \frac 12(1\mp \frac 1{%
\sqrt{3}})|010\rangle +\frac 1{\sqrt{3}}|100\rangle \mp \frac 12(1\pm \frac 1%
{\sqrt{3}})|001\rangle );  \nonumber \\
E_{7,8} &=&J+\frac{J_z}2+B\pm \eta _{+}:|\Phi _{7,8}\rangle =\frac 1{\sqrt{%
2\eta _{+}[\eta _{+}\pm (J_z+2B-J)]}}[(J_z+2B-J\pm \eta _{+})|111\rangle
+J\gamma \sum_{n=0}^2\Upsilon ^n|010\rangle ].
\end{eqnarray}
where $\eta _{\pm }=\sqrt{(J_z-J\pm 2B)^2+3(J\gamma )^2},\Upsilon $ is the
cyclic right shift operator \cite{wxg3}. The reduced density matrix of two
nearest-neighbor qubits in $N$ qubits system also has the form of Eq.(3).
Employing Eq.(4) and tracing on the basis of eigenstates shown in Eq. (7),
one can get the density matrix $\mu _1$, $\mu _2,w$, $z,v$, then further
obtain the concurrence. Here we do not write the expressions of $\lambda _i$
because it is very long. We will directly plot some curves to show the
effect of $J_z$ on enhancing entanglement.

Fig.4 show concurrence as a function of $B$ and $T$ with $\gamma =0.3$, $%
J_z=0.9$ and $J=1.0$ in three-qubit $XYZ$ Heisenberg chain. We see that with
the same $\gamma =0.3,$ the effect of partial anisotropy $\gamma $ make the
revival phenomenon more apparent than in two-qubit chain. When $B=4$ in Fig.
1, the largest critical temperature $T_c$ produced by $\gamma $ is about $%
1.0 $ (Fig.1a); due to the restrain of $J_z$ the maximum temperature only
caused by $\gamma $ is below 0.8(Fig.1b). However, in three-qubit system if $%
B=4$ with the same set of parameters, comparing Fig.1b with Fig.4, the
critical temperature $T_c$ in revival region almost equal to $1.8$. The
stronger effect of $\gamma $ implies that if we aim to obtain strong
entanglement we can decrease $\gamma $ properly and increase $J_z$,
otherwise increasing $\gamma $ can make the revival phenomenon more evident.
Of course, the coupling constant $J_z$ also increase magnetic field $B_c$
and expend the region of concurrence keeping constant in terms of $B$ and $T$
as it do in two-qubit (for the limited of the page,we do not plotted here) .

For $T=0.6$, Fig.5 show concurrence as function of $B$ and $J_z$. There is
no entanglement for $B=0$, which corresponds with Fig.4. If $J_z$ is below a
certain value, in case of Fig. 5 the value \medskip is about $0.2$, the
entanglement appears in one area corresponding to the ''revival''\cite{GL}
on condition that the magnetic field is larger than a certain value, and the
certain value of $B$ is increased with the enhanced of $J_z$. But, if $J_z$
is larger than $0.2$, there are two areas appearing entanglement, and the
entanglement appearing in the lower range of $B$ can be much stronger than
that in higher magnetic field. In the lower range of $B$, for a certain $B$,
the large $J_z$ the large concurrence. Thus, in the $N$-qubits $XYZ$ system,
for a fixed $T$, one can obtain a robust entanglement by controlling $B$ and 
$J_z$.

\section{Conclusion}

The thermal entanglement in anisotropic $XYZ$ Heisenberg chain is
investigated. Through analyzing the two-qubit system, we find that with the
increasing of $J_z,$ the critical magnetic field $B_c$ is increased; the
coupling along $Z$ not only improves the critical temperature $T_c$, but
also enhances the entanglement for certain fixed $B$. We also analyze the
entanglement between two nearest neighbors in three qubits and find that the
effect of partial anisotropy is more evident than it do in two-qubit system.
The pairwise entanglement exhibit a interesting phenomenon. For certain
fixed $B$ , if the coupling constant $J_z$ is small, the pairwise
entanglement only exists in relative strong magnetic field $B$ and the
entanglement is weak. By increasing $J_z,$ in lower range of $B$ , one can
obtain a strong entanglement. Therefore, interaction constant of the
z-component of two neighboring spins $J_z$ play important role in enhancing
entanglement and in improving the critical temperature.

\smallskip This work was supported by Ministry of Science and Technology of
China under Grant No.2100CCA00700\smallskip

The captions of the figure:

Fig. 1 Concurrence in two-qubit Heisenberg $XYZ$ chain is plotted vs $T$ and 
$B$, where (a): $J_z=0$, (b): $J_z=0.9$. For all plotted $J=1.0$, $\gamma
=0.3$.

Fig. 2 Concurrence in two-qubit Heisenberg $XYZ$ chain vs $B$ at zero
temperature for various values of $J_z$ with $\gamma =0.3$ and $J=1.0$. From
left to right $J_z$ equal to 0, 0.5, 0.9, respectively.

Fig.3 Concurrence in two qubits Heisenberg $XYZ$ chain is plotted vs $T$ .
For all plotted $J=1.0$, $B=1.1$,$\gamma =0.3$.From top to bottom $J_z$
equal to 0.9, 0.5, 0.2, 0, respectively.

Fig.4 Pairwise entanglement in three-qubit Heisenberg $XYZ$ chain is plotted
as a function of $T$ and $B$, where $\gamma =0.3$, $J=1.0$, $J_z=0.9$.

Fig.5 Pairwise entanglement is plotted as a function of $B$ and $J_z$, where 
$T=0.6$, $J=1.0$, $\gamma =0.3$ .

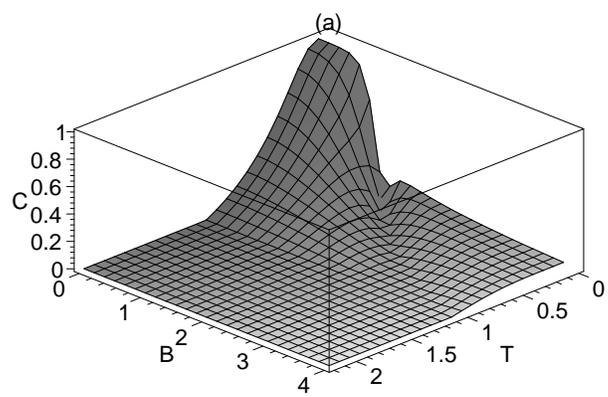
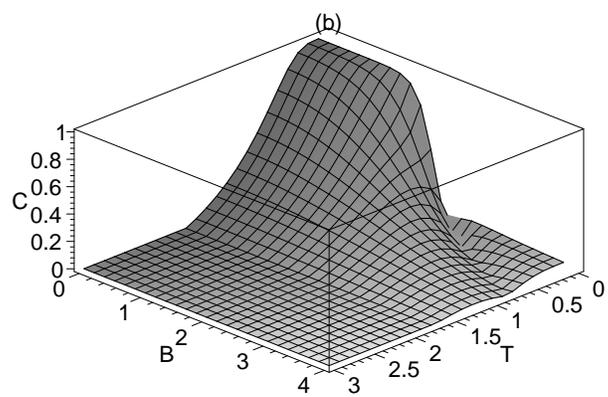
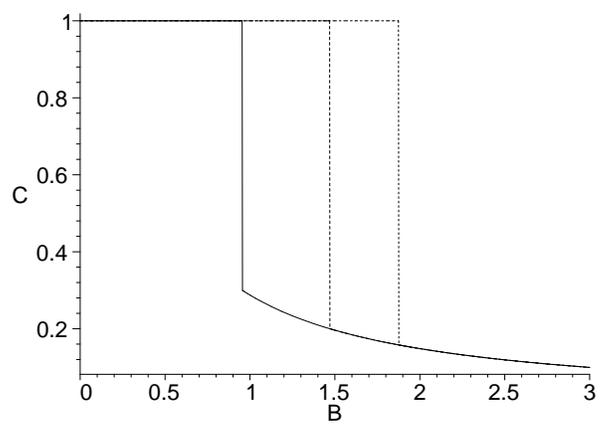
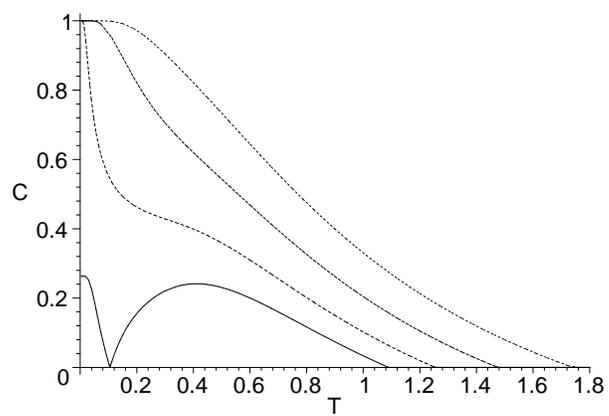
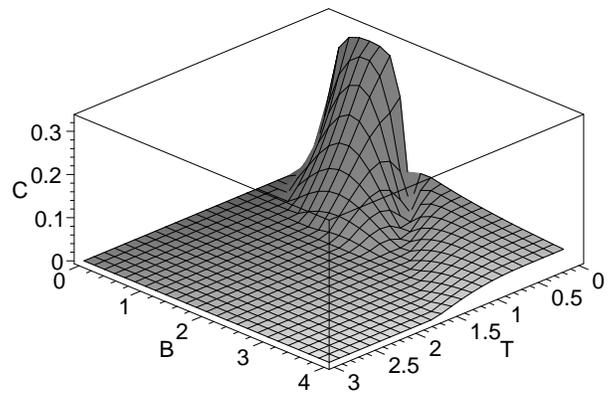
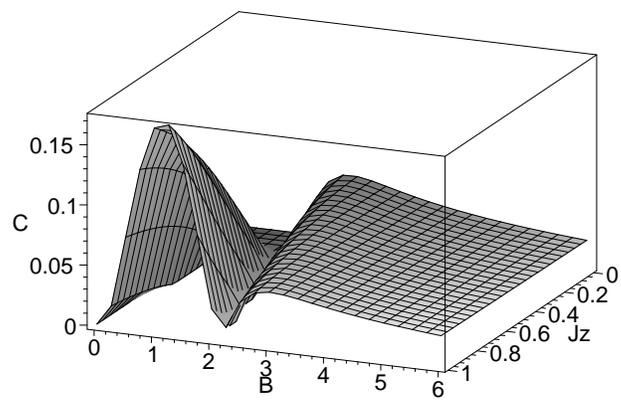
\end{document}